\documentclass[english]{article}
\usepackage[T1]{fontenc}
\usepackage{textcomp}
\usepackage[latin9]{inputenc}
\usepackage{color}
\usepackage{float}
\usepackage{graphicx}
\usepackage{esint}

\makeatletter
\newcommand{\lyxaddress}[1]{
	\par {\raggedright #1
	\vspace{1.4em}
	\noindent\par}
}

\usepackage{babel}
\usepackage[superscript,biblabel]{cite}

\makeatother

\usepackage{babel}
\begin{document}
\title{Microscopic Structure of Aqueous Alkylamine mixtures: a Computer Simulation
Study}
\author{Martina Požar$^{1}$, Lena Friedrich$^{2}$, Bernarda Lovrin\v{c}evi\'{c}$^{1}$,
\\
 Michael Paulus$^{2}$, Christian Sternemann$^{2}$ and Aurélien Perera$^{3}$
\thanks{aup@lptmc.jussieu.fr}}
\maketitle

\lyxaddress{$^{1}$Faculty of Science, University of Split, Ru\dj era Boškovi\'{c}a
33, 21000 Split, Croatia}

\lyxaddress{$^{2}$Fakultät Physik/DELTA, Technische Universität Dortmund, D-44221
Dortmund, Germany}

\lyxaddress{$^{3}$Laboratoire de Physique Théorique de la Matière Condensée
(UMR CNRS 7600), Sorbonne Université, 4 Place Jussieu, F75252, Paris
cedex 05, France.}
\begin{abstract}
Aqueous alkylamine mixtures are studied by computer simulations in
order to understand the microscopic origin of the water rich side
prominent x-ray scattering pre-peaks reported in a recent study. These
pre-peaks are puzzling in view of the apparently contradicting facts
that neat amines show pre-peaks much weaker than neat alkanols, while
water-rich aqueous alcohols do not. These observations can be intuitively
rationalized by noting that the amine head group have two hydrogen
atoms when the hydroxyl group have only one, but they oppose the following
two facts: i) computer simulations show micro-heterogeneity for both
systems; ii) amines mix with water better than alcohols, both over
larger concentrations and alkyl tails lengths.

The study of the atom-atom pair correlation functions and related
structure factors allows to understand the microscopic molecular details.
The most interesting observation is that the amine head groups accumulate
preferentially at the surface of the water domains, and increasingly
better with longer alkyl tail, thus allowing to stabilize both the
water and alkylamine domains, hence avoiding macroscopic demixing,
except at high water concentrations when amines are scarce to achieve
efficient surface saturation. The amine domains appear as disordered
bilayers. Hence, aqueous amines are analogous to an inverse micelle
melt and as precursor micro-emulsion.

This stable micro-segregation produces large domain oscillations in
the long range part of the correlation functions, translating into
positive pre-peaks and negative anti-peaks in the related structure
factors, the latter which contribute destructively to produce the
prominent scattering pre-peak observed in the x-ray experiments.

The model dependence is shown to be quite important, both for water
and solute models. The CHARMM-AA model associated with the SPC/E model
seems to be a good compromise.
\end{abstract}

\section{Introduction\protect\label{sec:Introduction}}

Alkylamines are known to be miscible in water \cite{Amine_Miscibility_Stephenson1993}
both over a large range of concentrations and alkyl chain length.
This is in contrast with alkanols, which demix with water beyond 1-propanol
and over a much larger concentration range. Interestingly, alkylamines
beyond pentylamine show a lower critical solution temperature (LCST).
From the interpretation of Walker and Vause \cite{LCST_Walker_Vause1983},
this indicates strong water-amine dimer type pairing. Another important
experimental fact is the large X-ray scattering pre-peaks (SPP) recently
reported for several aqueous amine mixtures\cite{our_expt_amin}.
SPP are often interpreted as a signature of the existence of supramolecular
associations. Indeed, SPP are found in several systems such as neat
alcohols \cite{EXP_Scatt_Narten_EthMeth,EXP_Sarkar_Joarder_Methanol,EXP_Sarkar_Joarder_ethanol,EXP_Joarder_Alcohols,EXP_Scatt_Finns_Monools,EXP_SANS_Soper_pure_methanol,EXP_Sarkar_1propanol,EXP_Scatt_Matija_Monools,EXP_Silleren_Propanol,2020_Linear_alcohols,2021_Octanols}
and amines \cite{2024_Neat_Amines}, room temperature ionic liquids,
both neat \cite{Ionic_1,Ionic_2,Ionic_3} and in aqueous mixing \cite{Ionic_water_Salma2016,Ionic_water_Salma2017}.
SPP are equally found in some water-poor aqueous alcohol mixtures,
such as aqueous octanol mixtures\cite{AUP-aqOct} . However, because
of the water strong tendency to self-Hbond, one expects small water
clusters in water-poor mixtures, hence such SPP are naturally expected.
Conversely, SPP from the water rich side are not obvious to explain
because of the complex interplay between the different Hbonding species
in presence.

Interestingly, SPP are preferentially found in soft matter systems
\cite{Soft_matter_PP_Paturej2022} and particularly micro-emulsions
\cite{EXP_Microemulsions1,EXP_Microemulsions2}. In what concerns
us, the SPP in aqueous amines appear to be inconsistent with the following
facts: i) neat alcohols have a much better defined SPP than neat amines;
ii) water-rich aqueous mixtures of simple alkanols do not show SPP,
rather they tend to show large $k=0$ concentration fluctuation peaks
\cite{EXP_Scatt_WaterEth_Nishikawa,EXP_Scatt_WaterAlc_Takamuku,EXP_Wat1prop},
usually a precursor signature of demixing \cite{AUP_Alcohol_Water_KBI2006}.
The difference between $k=0$ peaks and SPP is an important point
related to the difference between concentration fluctuations and micro-heterogeneity,
which we have addressed in previous works \cite{2015_PCCP_benzMH,AupPAC,Book_Smith2013fluctuation,AUP_acetone_2,AUP_dioxane_Kolarikova2024},
and which we will examine again in the later sections. These questions
are best answered through molecular models and computer simulations.

The physics behind SPP is quite simple from the point of view of Coulomb
association and charge order in molecular models: the molecular sites
of opposite charges tend to associate one another, hence segregating
the uncharged (or weakly charged) ones. This is for instance the case
in neat alcohol, where the OH hydroxyl head groups self-associate
in chain patterns \cite{AUP_Neat_Alcohols_JCP,AUP_Neat_alcohols_PRE,SIM_Bako_Neat_Meth_Clusters,SIM_Bako_methanol_EXP,SIM_Finci2_linear_alcohols,2020_Linear_alcohols,2021_Octanols}.
In the neat amines, we have shown that the Coulomb association is
hindered by the dual hydrogen atoms of the amine head group, leading
to weaker SPP \cite{2024_Neat_Amines}. In aqueous mixing conditions,
water enters the competition of Coulomb associations. It is intuitively
easy to imagine that the dual hydrogens of the amine groups are now
an advantage to associate with water, supporting the dimer picture
of Walker and Vause. However, it does not explain the prominent SPP
in a simple intuitive way.

The purpose of the present work is to examine the micro-structure
of aqueous amines for several amines ranging from propylamine to octylamine,
and for different water and amine models. All models show that aqueous
amine mixtures have considerable micro-heterogeneity, somewhat similar
to other aqueous mixtures, but which do not necessarily show SPP,
hence indicating that SPP and micro-segregation are not necessarily
equivalent. This can be explained only through the analysis of atom-atom
pair correlations, both in direct and reciprocal space. 

To our knowledge, there are very few simulation studies focusing purely
on the structuring in aqueous amines. Kusalik and co-authors published
a paper on the local structure in neat methylamine and methylamine-water
mixtures \cite{kusalik2000local}. The work of Lachet and colleagues
centered on the development of force fields of primary, secondary,
and tertiary amines \cite{Lachet_amine_FF_1,Lachet_FF_amines_2},
which were then put to use in studies about equilibrium and transport
properties of neat amines \cite{Orozco2014equilibrium_amines} and
gas solubility in amines \cite{Orozco2016physical_absorption_amines}. 

The remainder of this paper is as follows. In the next section we
study theoretical and methodological aspects (simulations). In the
third section we discuss the special type of micro-segregation observed
in simulations of aqueous amines. In the fourth section we study how
correlation functions help understand the micro-structure of these
mixtures. A discussion and conclusion sections close this paper.

\section{Theoretical and technical considerations\protect\label{sec:Methods}}

\subsection{Scattering intensity and density pair correlation functions}

The experimental x-ray scattering intensity for a binary mixture can
be readily obtained from the total atom-atom structure factors $S_{i_{a}j_{b}}^{(T)}(k)$,
where $a$, $b$ are molecular species indices and $i_{a}$, $j_{b}$
are atom index, through the Debye formula \cite{Debye1,Debye2}:

\begin{equation}
I(k)=r_{0}^{2}\rho\sum_{ab}\sqrt{x_{a}x_{b}}\sum_{ij}f_{i_{a}}(k)f_{j_{b}}(k)S_{i_{a}j_{b}}^{(T)}(k)\label{Ik}
\end{equation}
where the $S_{ij}^{(T)}(k)$ are defined as:
\begin{equation}
S_{i_{a}j_{b}}^{(T)}(k)=w_{i_{a}j_{b}}(k)\delta_{ab}+\rho H_{i_{a}j_{b}}(k)\label{ST}
\end{equation}
where the $w_{i_{a}j_{b}}(k)$ are the intra-molecular structure factors
(which imposes $a=b$, since these exists only within the same molecule,
hence the Kronecker symbol $\delta_{ab}$), $\rho=N/V$ is the total
number density (where $N=N_{a}+N_{b}=N_{1}+N_{2}$ is the total number
of molecules and $V$ the volume), $x_{a}=N_{a}/N$ is the mole fraction
of species $a$ (with notation $a=1,2$ for a binary mixture), and
the $H_{i_{a}j_{b}}(k)$ related to the Fourier transform of the atom-atom
intermolecular pair correlation function $g_{i_{a}j_{b}}(r)$ 
\begin{equation}
H_{i_{a}j_{b}}(k)=\sqrt{x_{a}x_{b}}\int d\vec{r}\left[g_{i_{a}j_{b}}(r)-1\right]\exp(i\vec{k}\cdot\vec{r})\label{Hk}
\end{equation}
In addition to $S_{i_{a}j_{b}}^{(T)}(k)$, we introduce the atom-atom
structure factor void of the intra-molecular contributions (the self
structure factors - see next sub-section)
\begin{equation}
S_{i_{a}j_{b}}(k)=\delta_{i_{a}j_{b}}+H_{i_{a}j_{b}}(k)\label{S(k)}
\end{equation}
which are the quantities which will be discussed in this paper.

The functions $g_{i_{a}j_{b}}(r)$ are calculated directly from the
GROMACS trajectory files (using the gmx rdf program), while the intra-molecular
parts $w_{i_{a}j_{a}}(r)$ are calculated as described in Ref. \cite{2020_Linear_alcohols}
by sampling the mean atom-atom distance histograms within each molecules
in several configurations\textcolor{blue}{.}

\subsection{Duality of concentration fluctuation and micro-heterogeneity\protect\label{subsec:Duality-of-concentration}}

The functions $g_{i_{a}j_{b}}(r)$ can be seen as atom pair \emph{distribution}
functions, but they are also defined as pair \emph{correlation} functions
in a statistical sense.

In the first case, the $g_{i_{a}j_{b}}(r)$ can be evaluated through
the histogram of the atom-atom spatial distributions in the molecular
configurations 
\begin{equation}
g_{i_{a}j_{b}}(r)=\frac{H_{i_{a}j_{b}}(r,\delta r)}{N_{i}^{2}\delta V(r,\delta r)}\label{eg-hist}
\end{equation}
where $H_{i_{a}j_{b}}(r,\delta r$) is the number of atoms $j_{b}$
in a spherical shell of radii $r$ and thickness $\delta r$, centered
on atom $i_{a}$, $N_{i}$ is the number of molecules of species $i$
in a volume V, and $\delta V(r,\delta r)=(4\pi r^{2}\delta r)/V$
is the normalized volume of the shell. This is done through the GROMACS
program.

In the second case, the $g_{i_{a}j_{b}}(r)$ are related \cite{Textbook_Hansen_McDonald}
to the second statistical moment of the random variable $\rho_{i_{a}}(\vec{r})$
, which is the microscopic atom density of atom $i_{a}$ defined as:
\begin{equation}
\rho_{i_{a}}(\vec{r})=\sum_{n_{a}}\delta(\vec{r}-\vec{r}_{n_{a}})\label{rho_ia-rando}
\end{equation}
The first moment is the one-body function $\rho_{i_{a}}^{(1)}(\vec{r})$,
which, in the absence of any external field, is just the number density
$\rho_{i_{a}}$ of atom $i_{a}$:

\begin{equation}
\rho_{i_{a}}^{(1)}(\vec{r})=<\rho_{i_{a}}(\vec{r})>=\frac{N_{i_{a}}}{V}=\rho_{i_{a}}\label{rho_ia}
\end{equation}
where $N_{i_{a}}$ is the number of atoms of type $i_{a}$, and <>
designates a statistical ensemble average.

The second moment is the two-body function function $\rho_{i_{a}j_{b}}^{(2)}(\vec{r}_{1},\vec{r}_{2})$
defined as
\begin{equation}
\rho_{i_{a}j_{b}}^{(2)}(\vec{r}_{1},\vec{r}_{2})=<\rho_{i_{a}}(\vec{r}_{1})\rho_{j_{b}}(\vec{r}_{2})>\label{rho_2}
\end{equation}
This function is related to the static van Hove function \cite{Textbook_Hansen_McDonald}
and can be split into a self part ($i_{a}=i_{b}$) and a distinct
part ($i_{a}\neq i_{b}$) by using Eq.(\ref{rho_ia-rando}) and separating
the two types of index contributions. The self part contributes to
the intra-molecular function \cite{2024_DYNA} and the distinct part.

The pair correlation function $g_{i_{a}j_{b}}(\vec{r}_{1},\vec{r}_{2})$
can be introduced from the distinct part $\rho_{D;i_{a}j_{b}}^{(2)}(\vec{r}_{1},\vec{r}_{2})$
through the obvious meaning of what a correlation is:
\begin{equation}
\rho_{D;i_{a}j_{b}}^{(2)}(\vec{r}_{1},\vec{r}_{2})=\rho_{i_{a}}^{(1)}(\vec{r}_{1})\rho_{j_{b}}^{(1)}(\vec{r}_{2})g_{i_{a}j_{b}}(\vec{r}_{1},\vec{r}_{2})\label{g(r)}
\end{equation}
Indeed, at large separations, this function should obey the relation
\begin{equation}
\lim_{|\vec{r}_{2}-\vec{r}_{1}|\rightarrow\infty}\rho_{D;i_{a}j_{b}}^{(2)}(\vec{r}_{1},\vec{r}_{2})=\rho_{i_{a}}^{(1)}(\vec{r}_{1})\rho_{j_{b}}^{(1)}(\vec{r}_{2})\label{lim_rho2}
\end{equation}
thus implying the well know limit for the correlation term
\begin{equation}
\lim_{|\vec{r}_{2}-\vec{r}_{1}|\rightarrow\infty}g_{i_{a}j_{b}}(\vec{r}_{1},\vec{r}_{2})=1\label{lim_gr}
\end{equation}
In the absence of an external field, the system is uniform and isotropic,
thus this function depends only on the module of the relative distance
$r=|\vec{r}_{2}-\vec{r}_{1}|$, and the correlation function naturally
reduces to $g_{i_{a}j_{b}}(r)$ which can be written from Eqs.(\ref{rho_ia},\ref{g(r)})
as
\begin{equation}
g_{i_{a}j_{b}}(r)=\frac{1}{\rho_{i_{a}}\rho_{i_{b}}}\rho_{D;i_{a}j_{b}}^{(2)}(r)\label{g(r)_iso}
\end{equation}
Equation (\ref{rho_2}) shows an important feature, that the pair
correlation function is related to the fluctuations of the microscopic
order parameters $\rho_{i_{a}}(\vec{r}).$ In that sense, they monitor
the local and global stability of the system \cite{Textbook_Chaikin_Lubensky}.
This is most efficiently seen through the small-$k$ behaviour of
the structure factors $S_{i_{a}j_{b}}(k)$ introduced in Eq.(\ref{S(k)}),
the $k=0$ values of which are related to the isothermal compressibility
\cite{Textbook_Hansen_McDonald,KBI_Seminal_Kirkwood1951,2022_Camel_back}.
The Kirkwood-Buff theory shows that the concentration fluctuations
in the system are related to the integrals of the pair correlation
functions.

What is generally observed is the following. When a mixture tends
to phase separate, then the concentration fluctuations are large,
their correlation range increases and this translates through Eqs.(\ref{Hk},\ref{S(k)})
into a $k=0$ growth of a peak in the structure factors. However,
when micro-heterogeneity and domain formation occurs, then it is a
positive small-k peak that appear for like species, usually accompanied
by a negative cross correlation anti-peak for unlike species \cite{AUP_ionic,Triolo_nanosegregation_2007,Ionic_1}.
In some systems, such as alcohol/ionic-liquid mixtures \cite{AUP_Charge_ordering_prepeak_neat_alc},
both phenomena are observed, which means that stable micro-segregation
can transform into macroscopic phase separation when parameters such
as concentration or temperature are tuned. The $k=0$ peak could be
considered as a packing of molecules in growing cluster, hence a classical
boson analogy. Conversely, the stable separation of antagonist nano-domains
could be considered a a classical fermion analogy. We previously suggested
this analogy in the context of aqueous-dioxane mixtures\cite{AUP_dioxane_Kolarikova2024}.

\subsection{Computer simulation technical details\protect\label{subsec:Simulation}}

The results presented in this work were obtained by molecular dynamics
simulations in the GROMACS program package \cite{MD_Gromacs}. We
tested several force fields for amines in order to reproduce the experimental
results of aqueous amine mixtures. We first tested united-atom models
of amines Gromos 53a6 \cite{FF_Gromos53a6}, Gromos 54a7 \cite{FF_Gromos_54a7}
and OPLS-UA \cite{FF_OPLS_AA_organic_liquids_amines,FF_OPLS_hydrocabons_UA},
which failed to yield miscibility with water. After that, we turned
to CHARMM all-atom \cite{FF_Charmm1,FF_Charmm2,FF_Charmm3} and OPLS-AA
\cite{FF_OPLS_AA_organic_liquids_amines,FF_OPLS_Amine_final}, which
produced miscibility with the SPC/e water model \cite{FF_SPCe}, but
also with the TIP4P\_2005 model \cite{FF_TIP4P_2005}. 

All aqueous amine mixtures contained 16000 molecules, the starting
configurations of which were generated with Packmol \cite{MD_Packmol}.
After obtaining the initial configurations, we followed the same simulation
protocol: energy minimization and equilibration for 10 ns, followed
by production runs of 10 ns, during which at least 2000 configurations
were collected. All simulations were performed in the NpT ensemble,
with the pressure \textit{p} = 1 bar and the temperatures of \textit{T}
= 300 K and \textit{T} = 340 K. The temperature was maintained with
the v-rescale thermostat \cite{MD_thermo_Vrescale} and the pressure
with the Parrinello-Rahman barostat \cite{MD_barostat_Parrinello_Rahman_1,MD_barostat_Parrinello_Rahman_2}.
The temperature algorithm had a time constant of 0.2 ps and the pressure
algorithm was set at 2 ps. The short-range interactions were calculated
within the 1.5 nm cut-off radius, while the long-range electrostatics
were calculated with the PME (Partial Mesh Ewald) method \cite{MD_PME}.
The LINCS algorithm \cite{MD_constraint_LINCS} handled the constraints.

\section{Mixing with nano-domain formation}

The experimental phase diagram from Glinsky et al.\cite{Phase_diagram_Glinski1994}
is sketched in Fig.SI-1 of the SI document, for aqueous hexylamine
and octylamine. One can see that both phase diagrams have a UCST (upper
critical solution temperature) behaviour, which means the existence
of a supra-structural organisation\cite{LCST_Walker_Vause1983}.

\begin{figure}[H]
\centering
\includegraphics[scale=0.25]{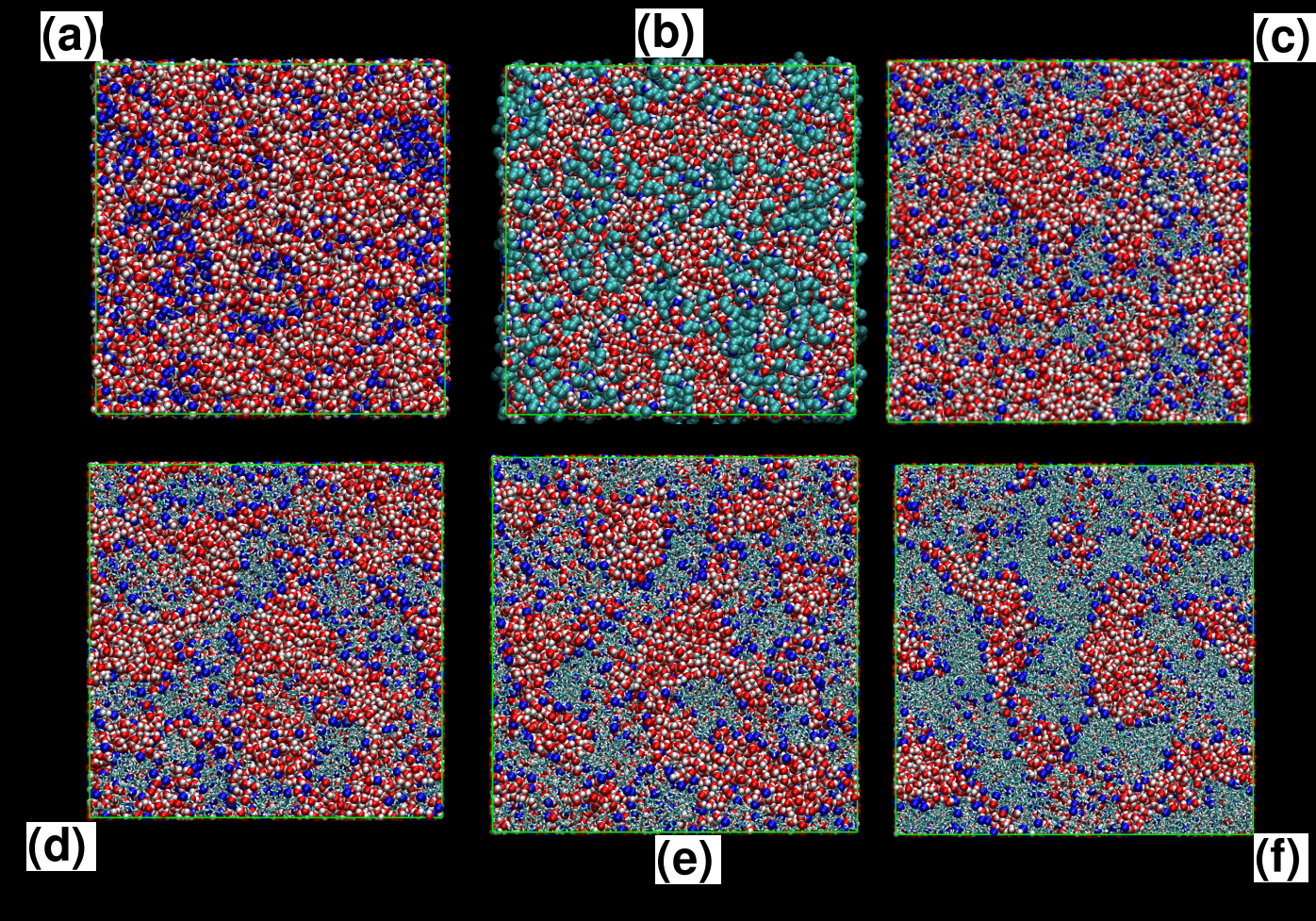}

\caption{Snapshots of one randomly chosen face of the simulation cubic box
for several aqueous mixtures of 20\% solutes. Panel (a) is for aqueous
ethanol (TraPPE model), panels (b) to (f) are for propylamine to octylamine
(skipping heptylamine), respectively. Water is shown with oxygen in
red and hydrogen in white. The amine nitrogen is highlighted in blue,
while all other amine atoms are shown in semi-transparent cyan. Ethanol
oxygen in (a) is shown in blue, and the propylamine alkyl tail in
(b) is shown in full atom representation (cyan). }

\label{Fig-snap1}
\end{figure}

Fig.\ref{Fig-snap1} illustrates what is perhaps the most astonishing
result of the computer simulations of aqueous alkylamines: that the
nitrogen atoms saturate the surface of the micro-segregated water
domains, hence contributing to stabilizing them. We believe that this
is the first time that this is shown for aqueous amines. All snapshots
show one face of the simulation box, chosen randomly, of a 20\% amine
mixture in water. The upper right panel shows aqueous ethanol (TraPPE
model) for comparison. The ethanol oxygen are shown in blue and it
can be seen that these are uniformly distributed throughout the segregated
ethanol domains. All other panels show alkylamines (CHARMM-AA model),
from propylamine in (b) up to octylamine in (f). While for propylamine
it is not obvious that the amine head groups are saturating the water
domains, this become quite clear starting from butylamine in panel
(c).

As the alkyl chains grow in length, the water domains shrink, but
the nitrogen atoms continue to saturate their surface, with very little
amine head inside the alkyl domains. This is consistent with the fact
that neat alkylamines do not form specific large amine head group
clusters \cite{2017_Propylamine1}, while neat alkanols tend to form
nice chain clusters \cite{AUP_Neat_Alcohols_JCP,AUP_Neat_alcohols_PRE}. 

\begin{figure}[H]
\centering
\includegraphics[scale=0.45]{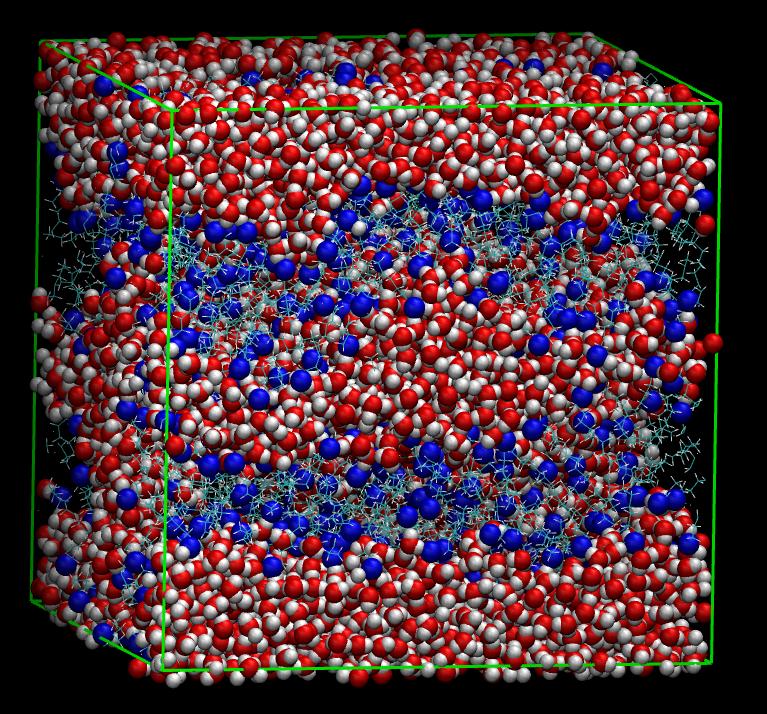}

\caption{Snapshot of a low temperature (10 °C) hexylamine-water bilayer system.
The atom color follow the same conventions as in Fig.\ref{Fig-snap1}.}

\label{Fig.snap2}
\end{figure}

The spacing between the water domains is more or less abou\textcolor{blue}{t}
two alkyl tail lengths, which would be consistent with a disordered
alkylamine bilayer system. Indeed, at low temperature, experiments
show that such bilayers are formed. Fig. \ref{Fig.snap2} shows such
a layer system for a 10\% hexylamine aqueous mixture at 10 °C, and
how such bilayers remain stable despite some fluctuations. The amine
bilayers are such that the amine head groups stick on both sides to
the upper and lower water layers.

As temperature is increased, these bilayers get disorganized and form
random micro-emulsions type patterns observed in Fig. \ref{Fig-snap1}.
At low amine densities, when the water domains tend to be large, there
are not enough amine head groups to stabilize them and the system
demixes \cite{Phase_diagram_Glinski1994}.

Interestingly, the low amine content demixing range increases with
temperature, which is a scenario consistent with the LCST found in
experiments \cite{Amine_Miscibility_Stephenson1993}. Indeed, as temperature
increases the water domain become more disordered, hence larger, which
makes it even more difficult for the amine head groups to cover them.

\begin{figure}[H]
\centering
\includegraphics[scale=0.3]{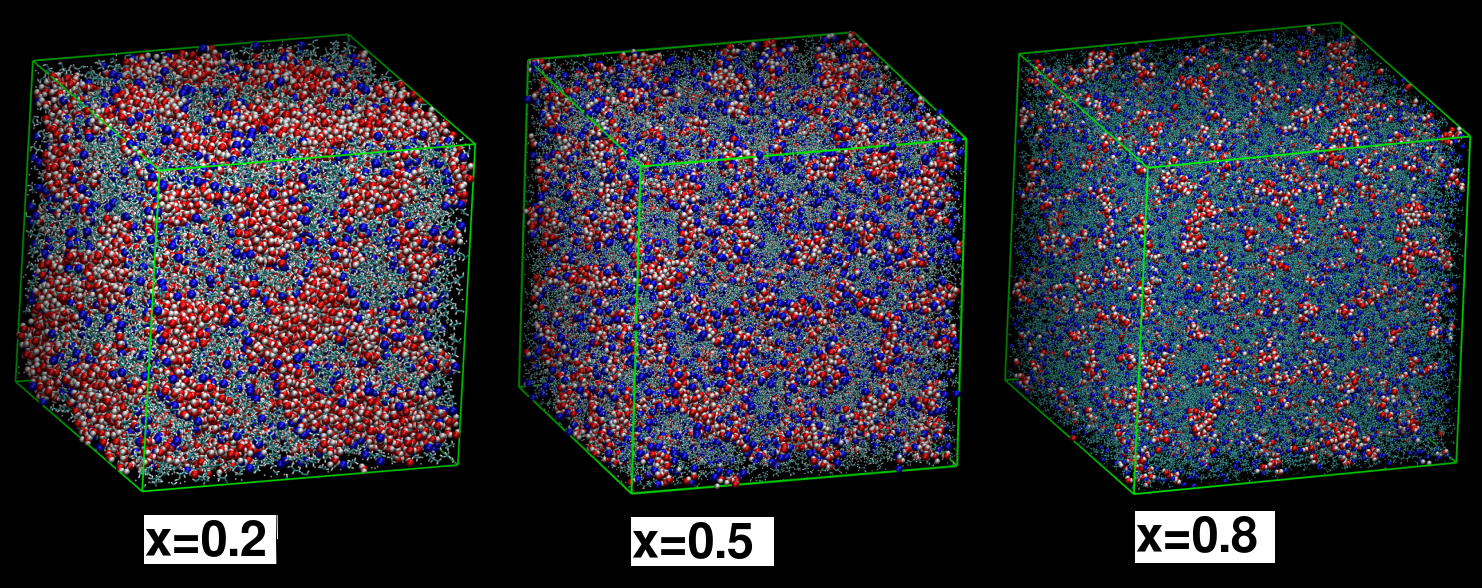}

\caption{Evolution of the domain structure with amine concentration $x$, for
the aqueous hexylamine systems. Atom coloring conventions as shown
in Fig.\ref{Fig-snap1} . The alkyl tails are represented in semi-transparent
modes, with color codes adapted to enhance domain size and structure
differences. }

\label{Fig.snap-conc}
\end{figure}

The amine concentration dependence of the microstructure is illustrated
in Fig. \ref{Fig.snap-conc} for the case of hexylamine. It is clearly
seen that the water domains shrink as the amine concentration increases.
But the nitrogen atom saturation of the domains is preserved, even
if for large amine concentrations many nitrogen atoms appear in the
bulk of the amine domains.

How does this microstructure translate into the various atom-atom
pair correlation functions? This is examined in the next section.

\section{Microscopic description from density pair correlations }

Since there are several atoms involved, it is necessary to sort out
which atom pairs are most relevant. Indeed, an all-atom description
of hexylamine requires 120 atom pairs. Below we examine selectively
the various combinations of correlations between the following atoms,
the water oxygen atom $O_{W}$, the amine nitrogen atom $N$ and the
first $C_{1}$ and last (tail) $C_{T}$ atoms. This way, we can have
a good idea of the disordered layer type correlations. Perhaps the
most interesting aspect is the long range domain correlations, which
correspond to the stable micro-heterogeneity observed in Fig. \ref{Fig-snap1}.
These are the correlations which are responsible for the SPP. We have
hinted to the existence of such features in aqueous 1-propanol mixtures
\cite{AupDomainOrdering}. But, the absence of SPP for these mixtures,
did allow to further develop the research along these lines. Aqueous
amine mixtures appear to be the missing link between solutions and
soft matter. It is therefore important to understand how one of the
most important observables from computer simulations behave for this
type of liquid mixtures. In particular, the relation to the dual fluctuation/micro-heterogeneity
discussed in Section \ref{subsec:Duality-of-concentration} needs
some support from the analysis of the correlation functions.

\subsection{Pair correlation functions of the H-bonding groups}

Because of the existence of dual type of distributions, namely the
usual first few neighbour molecular dispositions as well as long range
domain correlations, we will focus on both aspects. We will consider
how these uncommon second types of domain correlations settle as the
alkyl tails of the amines are made longer and longer, from propylamine
to octylamine. For this purpose, we focus on $T=300$ K mixtures with
amine mole fraction $x=0.2$. We will study charge-charge correlation
of the water and amine head groups, as well as those involving neutral
alkyl tail atoms.

\begin{figure}[H]
\centering
\includegraphics[scale=0.4]{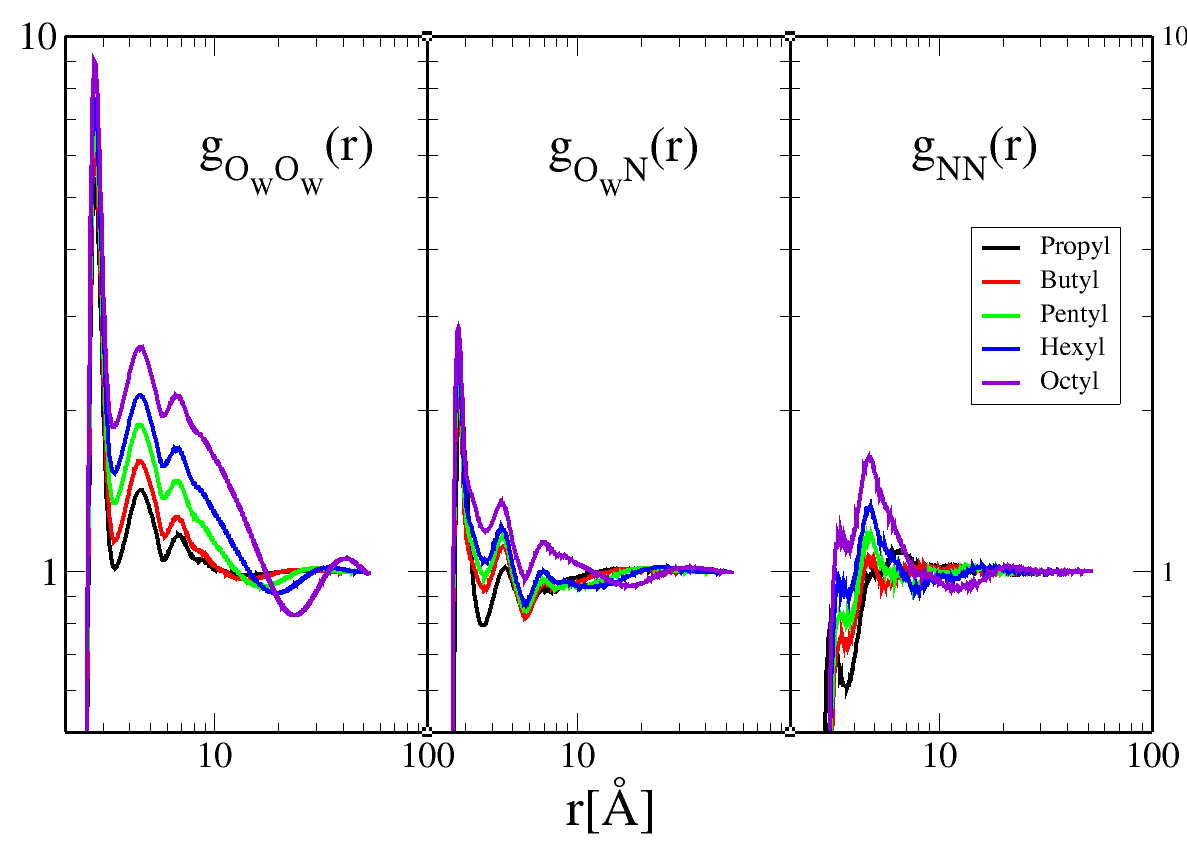}

\caption{Selected atom-atom correlation functions (in log/log) for various
alkylamine-water mixtures for amine mole fraction $x=0.2$. The domain
oscillations (magnified by the log-log plot) are quite visible through
the water-water correlation (left panel), specially for longer amines,
with domain size about 12nm. For comparison, the semi-log plots are
shown in the SI in Fig-SI-1.}

\label{Fig-gr-all}
\end{figure}

Fig. \ref{Fig-gr-all} shows the pair distributions from the three
principal Hbonding atoms pairs, namely oxygen-oxygen $g_{O_{W}O_{W}}(r)$
functions for water, nitrogen-nitrogen $g_{NN}(r)$ for amines, and
the cross correlations $g_{O_{W}N}(r)$. These are represented in
an unusual log-log plot (see Fig.SI-2 for a more conventional view),
in order to magnify at the same level both the large short range correlations
due to immediate Hbond neighbour interactions, and the long ranged
domain correlations.

The comparison of the left and right panels shows that the water-water
correlations are stronger than the solute amine-amine correlations,
which is also observed in alcohol-water mixtures \cite{AUP_Ethanol_JMolLiq,AUP_Asenbaum_ethanol}.
A striking feature is the high first peak of $g_{O_{W}O_{W}}(r)$
compared with the much smaller Hbond N-N peak at $r\approx4\mathring{A}$
in $g_{NN}(r)$, the latter which is even smaller than the broader
main peak at $r\approx4\mathring{A}$, thus indicating that nitrogen
atoms nearby to other nitrogen atoms are not always Hbonded, which
is not the case for water. The middle panel for $g_{O_{W}N}(r)$ confirms
that there are strongly Hbonded O-N pairs, even if the corresponding
correlations are about three times smaller than water-water Hbonds. 

\begin{figure}
\includegraphics[scale=0.4]{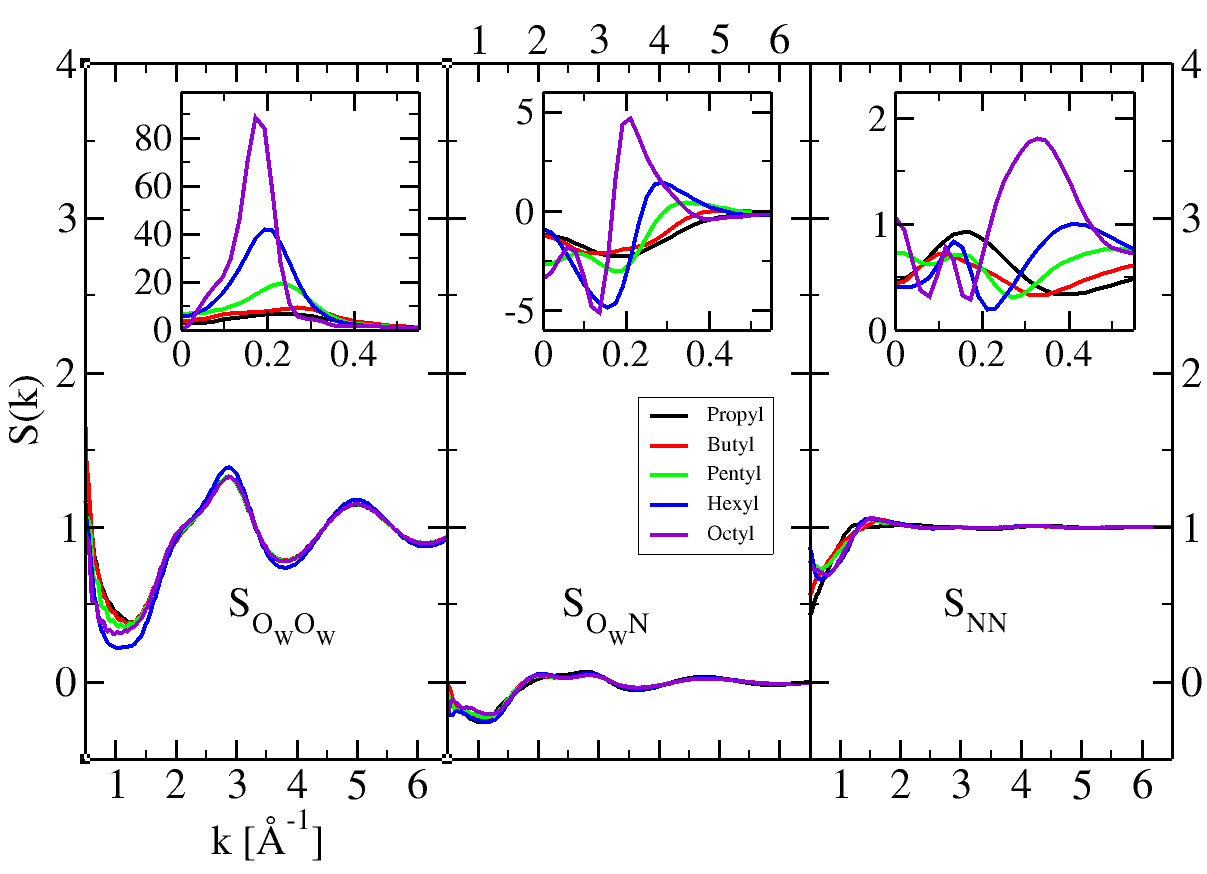}

\caption{Structure factors corresponding to the pair correlation functions
in Fig.\ref{Fig-gr-all}, with same color conventions.}

\label{FIg-Sk-ALL}
\end{figure}

The atom-atom structure factors corresponding to the pair correlation
functions above are shown in Fig.\ref{FIg-Sk-ALL}. Perhaps the most
spectacular feature is the fact that the water-water structure factors
for all solutes are nearly indistinguishable from that of neat water
in the first peak $k$-range and beyond $k>1.5\mathring{A}$. The
split feature is the same, with the H-bond peak at $k\approx2\mathring{A}$
and the contact peak at $k\approx3\mathring{A}$. This is an outstanding
feature because of the following two implications. First, it means
that water preserves its bulk type Hbonding even in presence of solutes
of different types. Second, one can therefore decouple the small-$k$
and in particular the pre-peak features from both the pair correlation
function and its structure factor in the following way
\begin{equation}
g_{WW}(r)=g_{W}(r)+g_{LR}(r)\label{gr-sep}
\end{equation}
\begin{equation}
S_{WW}(k)=S_{W}(k)+S_{PP}(k)\label{Sk-sep}
\end{equation}
where $g_{W}(r)$ and $S_{W}(r)$ are the neat water atom-atom pair
correlation and structure factor, respectively, for every atom pair
OO OH and HH, and $g_{LR}(r)$ is essentially the long range feature
(together with remaining small range details), and $S_{PP}(k)$ is
the corresponding Fourier transform. These two features, which are
important for self-assembly, will be studied in greater details in
a companion paper\cite{our_TS_amin}.

In contrast to the water-water correlation functions, the cross correlations
(middle panel) and solute-solute correlations show mixed contributions
that cannot be detached so clearly as for water, even though this
could be formally possible. The enormous difference in magnitude of
the pre-peaks between the water and the other contribution is an additional
signature of the importance of water preserving its bulk structure.
The pre-peaks are direct consequence of the domain oscillatory features
observed in Fig.\ref{Fig-gr-all} in the long range parts of the pair
correlation functions.

\begin{figure}
\centering
\includegraphics[scale=0.4]{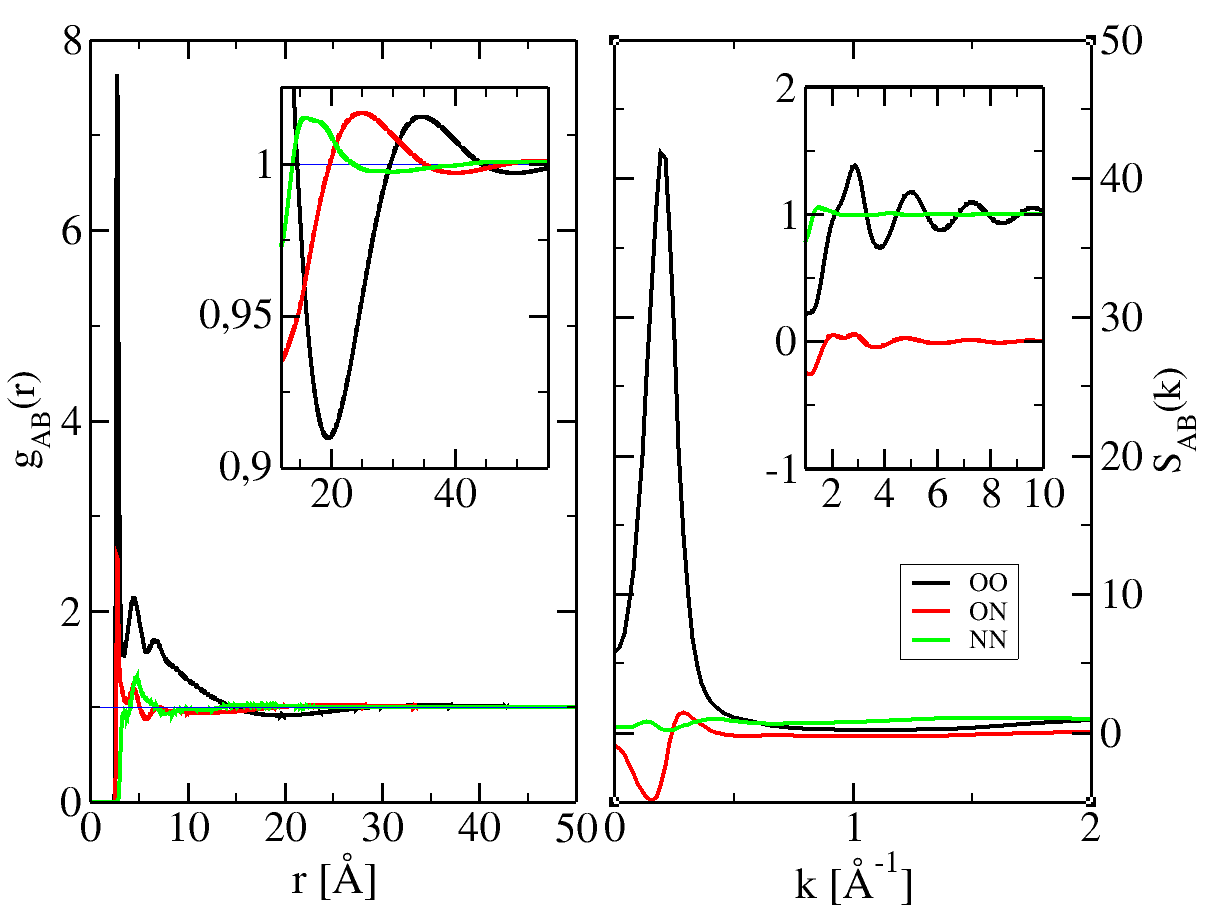}

\caption{Short range correlations versus domain correlations, illustrated for
the case of 20\% aqueous hexylamine mixtures, with correlations between
the water oxygen $O_{W}$ and amine nitrogen $N$ (see inset for color
conventions). Left panel shows the short range correlations in the
main panel and the long range domain oscillations in the inset. Right
panel shows small-$k$ SPP features in the main panel and larger $k$
features in the inset.}

\label{Fig-DomainCorr}
\end{figure}

Fig. \ref{Fig-DomainCorr} shows an important feature of the domain
correlations, which is the fact that the water-amine cross domain
correlations (red) are in phase opposition with the water-water (black)
and amine-amine (green) correlations. 

It is this important feature which leads to the cancellations of the
prominent partial species contributions to the radiation scattering,
leading themselves to a reduced but still prominent scattering pre-peak,
as will be discussed later in Section \ref{sec:X-ray-scattering}
.

\begin{figure}
\centering
\includegraphics[scale=0.4]{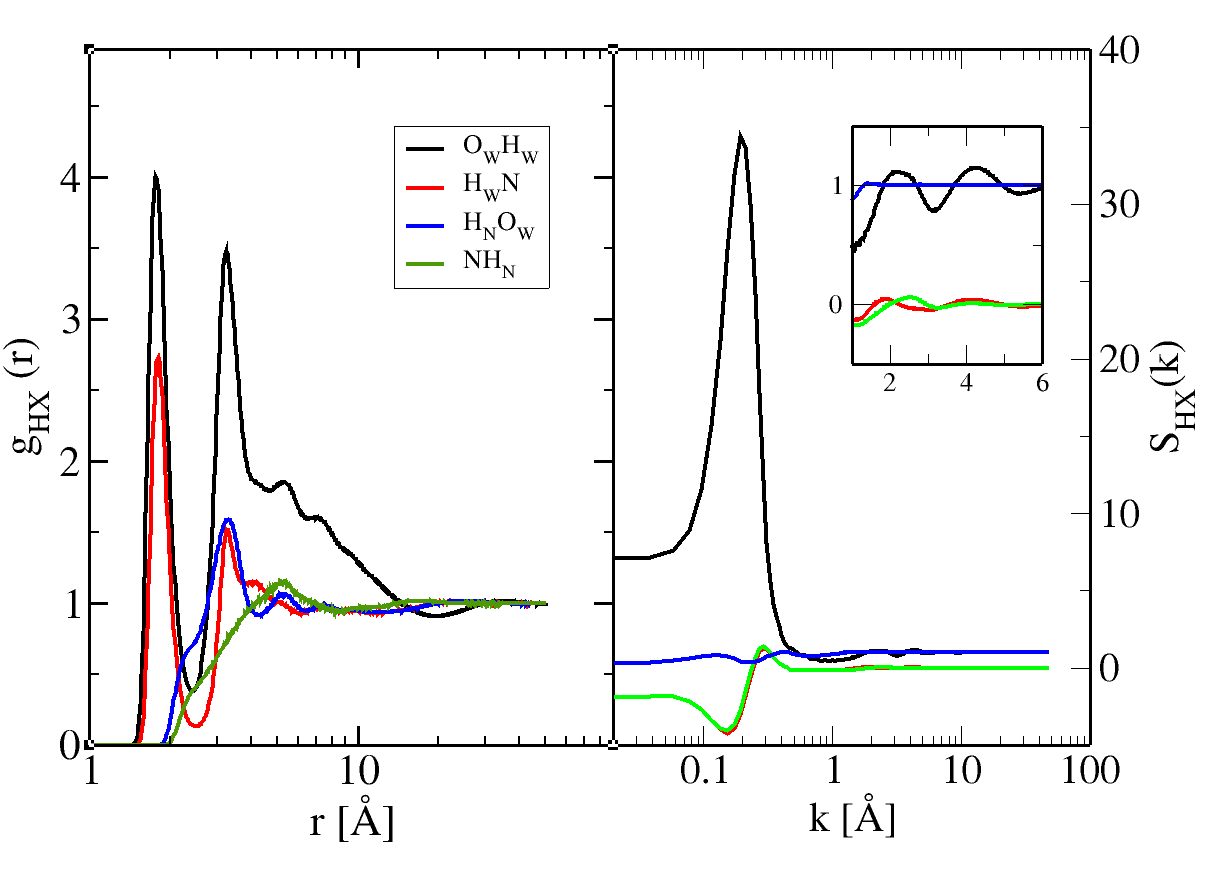}

\caption{Hydrogen atom correlations between water $H_{W}$ and amine $H_{N}$
atoms with water oxygen $O_{W}$ and amine nitrogen $N$ atoms. Left
panel shows the $g(r)$ and right panel the corresponding $S(k)$.
The inset shown a zoom around the main peak at $k\approx2-3\mathring{A}^{-1}$.}

\label{Fig-HX}
\end{figure}

It is equally important to study the Hbonding correlation between
water and the solutes. This is sketched in Fig. \ref{Fig-HX}.

The H atom correlations reported in Fig. \ref{Fig-HX} and corresponding
to various Hbonding patterns, illustrated in the case of aqueous 20\%
hexylamine, confirm the deductions above, that water-water correlations
(black) are the most prominent, followed by water-amine Hbonding (red).
Of the two remaining lesser Hbonding the amine-amine (blue) is nearly
featureless, witnessing very little Hbonding. It is remarkable that
the water-amine HN Hbonding is so much more important than the amine-water
HNO, indicating that the water-amine dimer through the O-H-N is quite
strong. This must be detected in Raman optical spectroscopy measurements,
although we are not aware of any such result.

All correlations increase in magnitude with longer alkyl chains, which
is a direct consequence of their hydrophobicity helping the mutual
confinement of the respective water and amine domains. This is the
origin of the micro-heterogeneity, equally observed in aqueous mixtures
of smaller alkanols \cite{MH_Soper_Dixit2002,MH_Guo2003,MH_Allison2005,MH_Patey_TBA_Wat_Gupta2012,2016_PCCP_MH_Versus_Clust},
but which is replaced by full demixing for longer alkanols. Instead,
the long range correlations seen in Fig.1 for all three correlation
functions, suggest that it is the mechanism behind these correlations
that prevents full demixing.

\subsection{Pair correlations involving alkyl tails}

In this section, we examine how the water oxygen $O$ and amine nitrogen
$N$ atoms correlate with the hydrophobic tails, typically the first
$C_{1}$ and last tail $C_{T}$ carbon atoms. Then, we also examine
how the tail carbon atoms correlate between themselves. 

\subsubsection{Charged groups correlations with the first carbon atom of the alkyl
tails}

\begin{figure}
\centering
\includegraphics[scale=0.4]{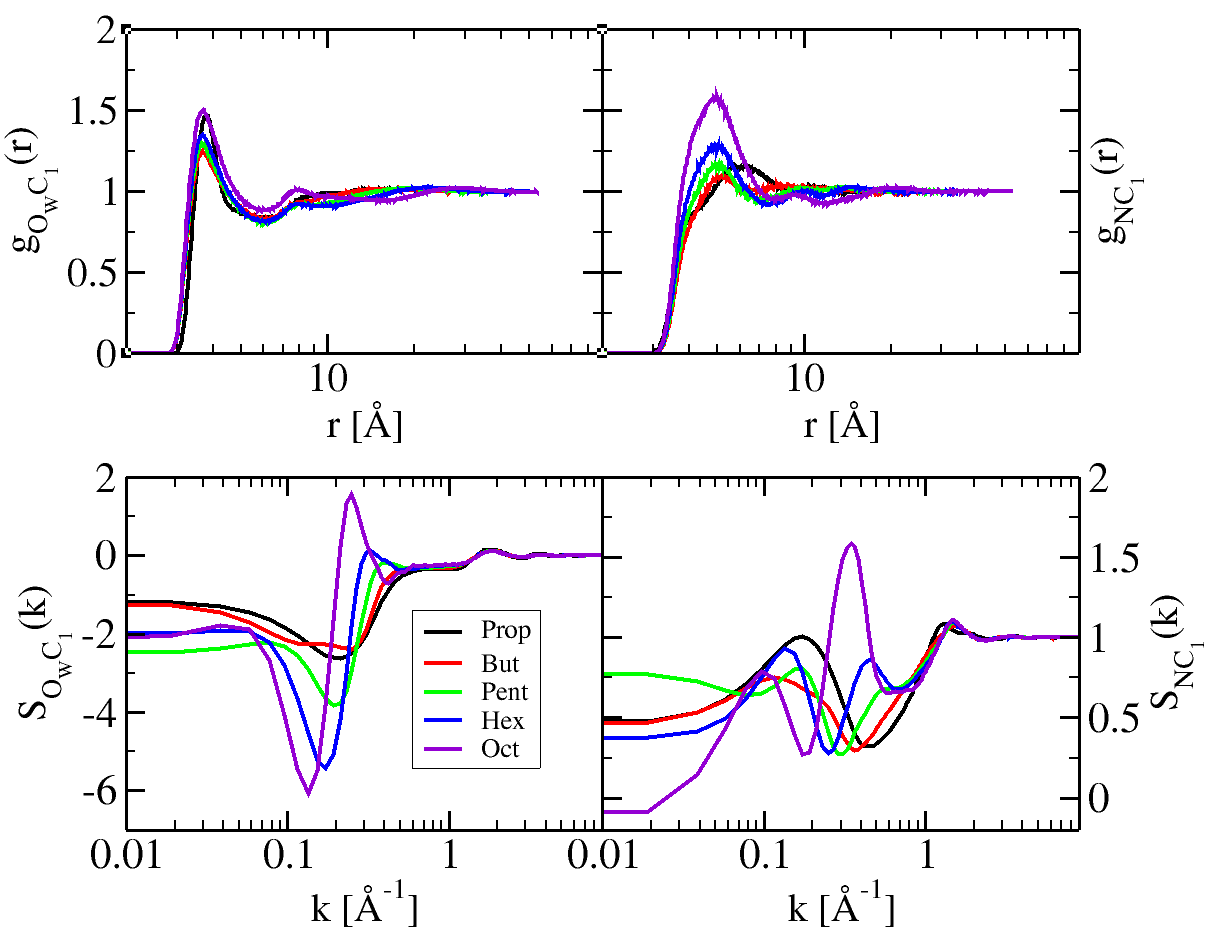}

\caption{Correlations between the charged atoms (water oxygen O and amine nitrogen
N) with the first carbon atom $C_{1}$ of the alkyl tails, from propylamine
to octylamine. The g(r) functions are shown in the upper panel, and
the S(k) functions are shown in the lower panels, left for water and
right for amine. The alkylamine color conventions for the curves are
shown in the legend box of the lower left panel.}

\label{FigC1-X}
\end{figure}

We first examine the water oxygen $O$ and amine nitrogen $N$ correlation
with the hydrophobic first alkylamine carbon atom $C_{1}$. In Fig.\ref{FigC1-X},
the upper left panel shows the $g_{O_{W}C_{1}}(r)$ functions for
the various alkylamines while the right panel shows the $g_{NC_{1}}(r)$.
What is quite apparent is that only the first neighbour correlations
are prominent, as in usual liquids, and the domain correlations are
quite small. While this is consistent for the $g_{NC_{1}}(r)$, in
view of the large domain correlations of the $g_{O_{W}N}(r)$ in Fig.\ref{Fig-gr-all},
one would have expected similar domain correlations in the $g_{O_{W}C_{1}}(r)$
since the $C_{1}$ atom is close to the amine head group. In fact,
the near similarities between $g_{O_{W}C_{1}}(r)$ and $g_{NC_{1}}(r)$
indicates that the water molecules form tight dimers with the $N$
atoms and are only indirectly correlated to the $C_{1}$ atoms. 

Nevertheless, we do observe depletion correlations of the second neighbours,
specially for longer amines (purple curve for octylamine for instance),
which is a trace of the interfaces formed by the water and amine head
group atoms. The structure factors $S_{O_{W}C_{1}}(k)$ and $S_{NC_{1}}(k)$
shown in the two lower panels confirms the observations made for the
respective $g(r)$ functions. For the amine-amine structure factors
we find weak pre-peak due to the Hbonding structures, as in the case
of the neat amines \cite{2024_Neat_Amines}. For the water-carbon
structure factor we find the typical charge order pre-peak/anti-peak
correlations. We note that both structure factors contribute only
weakly to the pre-peak, unlike those for the charged groups seen in
the insets Fig. \ref{FIg-Sk-ALL}. 

\subsubsection{Charged atom correlations with the last carbon atom of the alkyl
tails}

\begin{figure}
\centering
\includegraphics[scale=0.4]{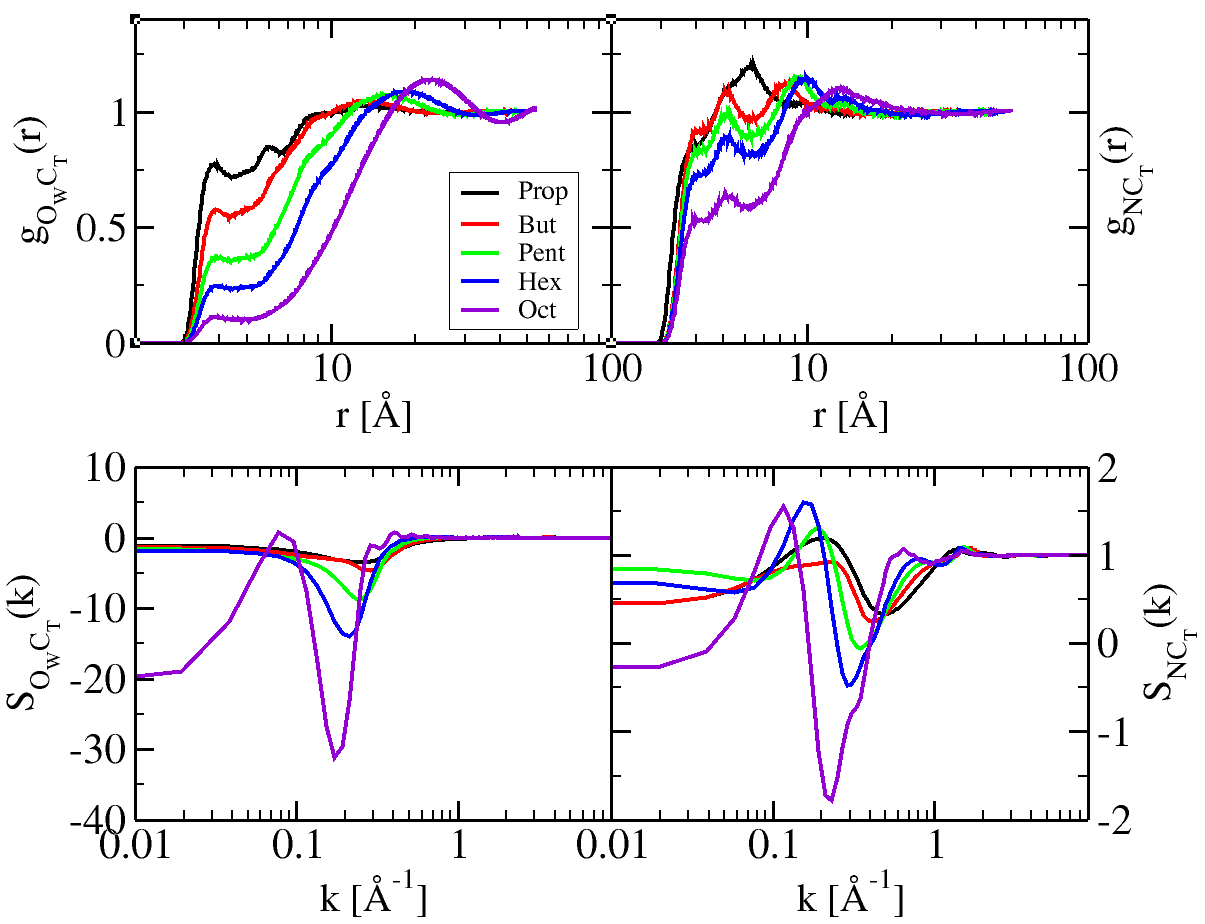}

\caption{Correlations between the charged atoms (water oxygen O and amine nitrogen
N) with the last carbon atom $C_{T}$ of the alkyl tails, from propylamine
to octylamine. The g(r) functions are shown in the upper panel, and
the $S(k)$ functions are shown in the lower panels, left for water
and right for amine. The alkylamine color conventions for the curves
are shown in the legend box of the upper left panel.}

\label{FigCt-X}
\end{figure}

We expect anti-correlation between the charged groups and the last
alkyl tail carbon atom, because of the hydrophobicity between the
two segments, and this is what comes out Fig. \ref{FigCt-X}.

What is quite apparent in upper left panel is that the $g_{O_{W}C_{T}}(r)$
are dominated by the domain correlations (large oscillations) while
the neighbouring atom correlations are the small wiggles (same range
as in Fig. \ref{FigC1-X} for instance), and the latter which are
seen to be more prominent for smaller amines and much less as the
alkyl tail length increases. Conversely, the domain correlations are
seen to increase in the opposite direction. Both features make sense,
since we expect more domain correlations for larger alkylamines. The
lower left panel shows the corresponding structure factors $S_{O_{W}C_{1}}(k)$.
We see very clearly that the large tail oscillations contribute to
the larger negative anti-peaks, more so as the alkylamine tails size
increase. The negative anti-peak indicates that the water oxygen amine
carbon atoms are depleted and in phase opposition with the charge
atoms correlations of Fig. \ref{Fig-gr-all} and Fig. \ref{FIg-Sk-ALL}.

We now turn to similar charged/uncharged atom correlations but within
the alkylamine molecules. The right panels show the nitrogen/first
carbon correlations, $g_{NC_{1}}(r)$ in the upper one, and $S_{NC_{1}}(k)$
in the lower one. Here, we observe only neighbouring atom correlations,
as in any standard liquids, and the domain correlations are very small.
This is consistent with what we observed in a previous work for the
neat alkylamines \cite{2024_Neat_Amines}, namely that the negative
anti-peak correlations were quite weak when compared with neat alcohols.
Indeed, the lower panel shows smaller pre-peak amplitudes than that
in the left panel for the oxygen/carbon atoms (as witnessed by the
different verticals between the two lower panels).

\subsubsection{Carbon-carbon correlations}

\begin{figure}
\centering
\includegraphics[scale=0.4]{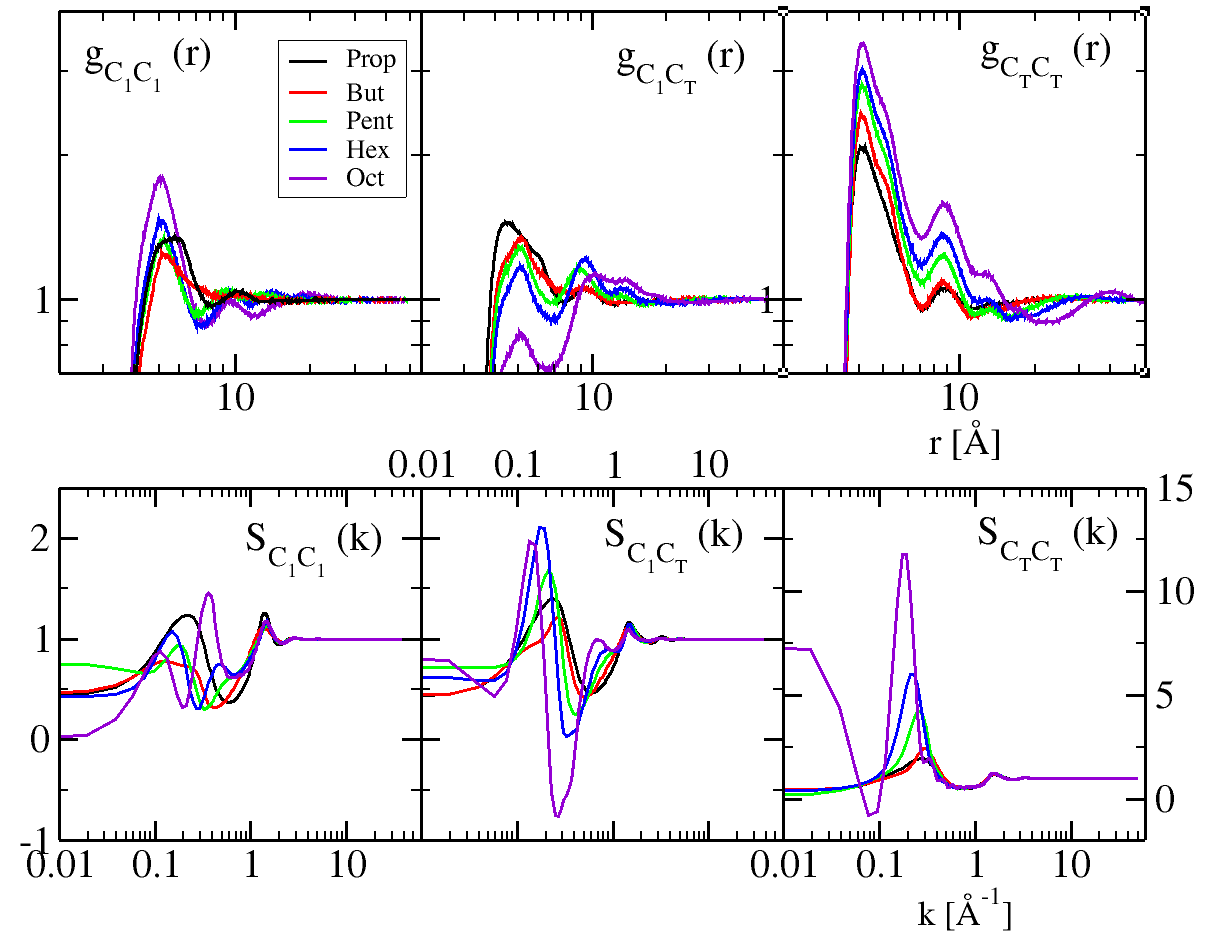}

\caption{Correlations combinations between the first $C_{1}$ and the last
$C_{T}$ carbon atoms of the alkyl tails, from propylamine to octylamine.
The $g(r)$ functions are shown in the upper panel, and the $S(k)$
functions are shown in the lower panels, left for water and right
for amine and cross correlations in the middle. The alkylamine color
conventions for the curves are shown in the legend box of the upper
left panel.}

\label{FigC1Ct}
\end{figure}

The correlations between the first $C_{1}$ and last $C_{T}$ carbons
are shown in Fig. \ref{FigC1Ct} and for different amines, in log
scale for the x-axis. The examination of the $g(r)$ functions in
the three upper panels reveals a striking observation, that the $C_{T}-C_{T}$
tail atom correlations dominates the other two by a factor 10 (y-axis
in log scale). This is a remarkable finding since it proves the layering
hypothesis of the alkyl domains. Indeed, if the micro-segregated alkyl
domains are constrained by the binding of the $O_{W}$ and $N$ atoms,
then it necessarily means that the bilayer structure must be more
or less preserved, even in disordered form. This enforces the correlations
between the last carbon atoms, and this is exactly what the $g_{C_{T}C_{T}}(r)$
show, and for all amines. 

This observation is further supported by the fact that the long range
correlations of the $g_{C_{T}C_{T}}(r)$ large domain correlations,
which is much less the case for the two other functions. Indeed, it
is only natural that the last carbon atom witnesses the segregation
of the alkyl tail domains from the water domains. The structure factors
in the lower row of Fig. \ref{FigC1Ct} confirm these findings. Indeed,
the amplitude of the pre-peaks for the tail atom $S_{C_{T}C_{T}}(k)$
are about 10 times more prominent than the two others, and more so
for the longer amines.

Carbon-hydrogen correlations of the alkyl tail atoms are shown in
the SI document, in Fig.SI-3 for the g(r) and Fig.SI-4 for the S(k).
These plots show clearly that the methyl groups closer to the amine
head group witness domain order more readily that the middle or tailing
groups. Also the carbon correlations dominate somewhat the hydrogen
and carbon-hydrogen correlation, which justifies to some extent the
united atom representation.

\section{Concentration and temperature dependence}

So far we have examined only the 20\% amine content between different
amines and for room temperature conditions. Since the phase diagrams
of the aqueous amines are available from experiments, it is interesting
to see how the simulations reproduce concentration and temperature
dependence's. We will illustrate this for the case of aqueous hexylamine
mixtures for the amine CHARMM-AA model.

\subsection{Concentration dependence}

\begin{figure}
\centering
\includegraphics[scale=0.4]{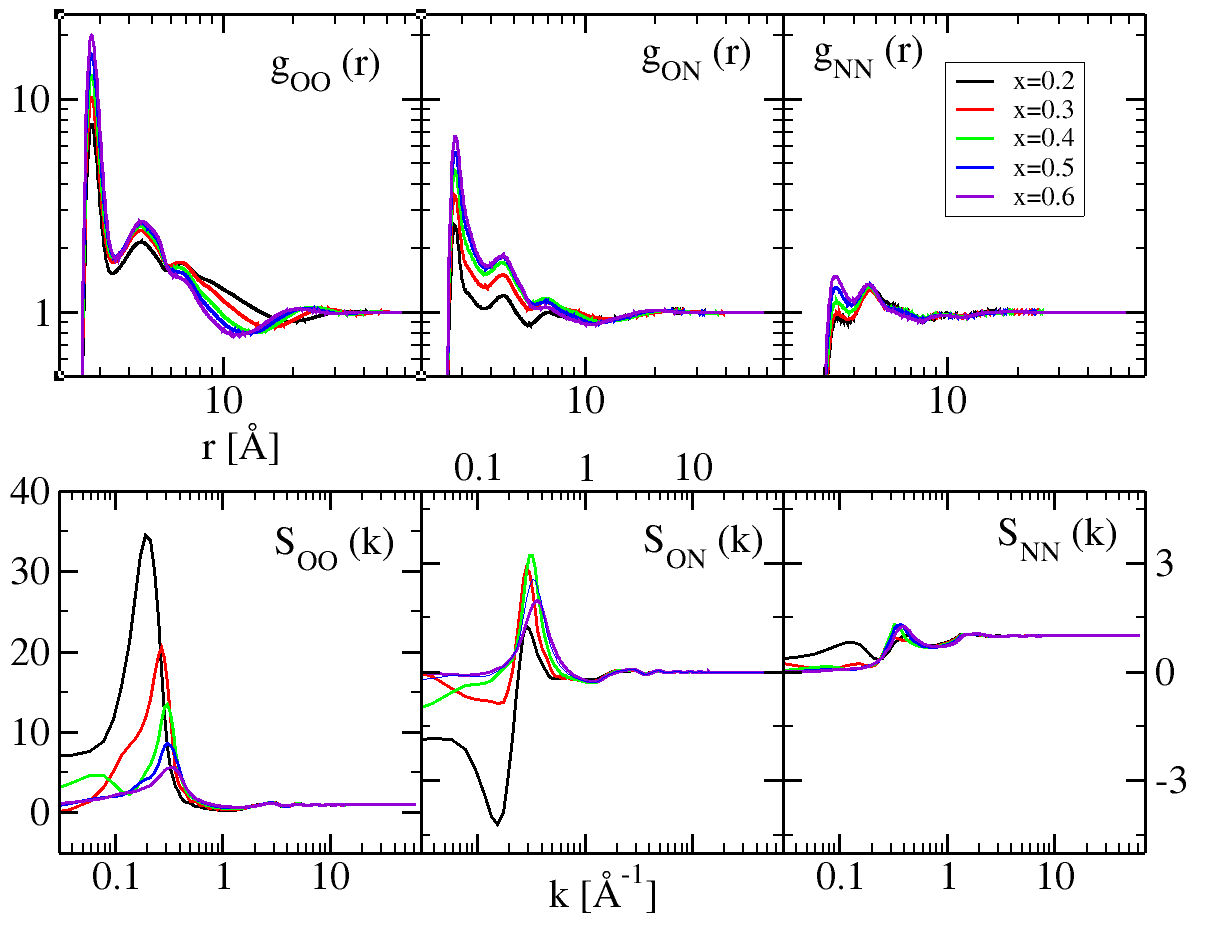}

\caption{Hexylamine (CHARMM-AA) concentration dependence of the charged atoms
$O_{W}$ and $N$ correlation functions (upper rows) and corresponding
structure factors (lower rows), shown in log-log scale. The vertical
scale for the middle and right lower panels is shown in the right
most vertical axis, since it is different from that of the lower left
panel.}

\label{Fig-Hex-density}
\end{figure}

Fig.\ref{Fig-Hex-density} shows the room temperature hexylamine (CHARMM-AA)
concentration dependence for the correlations between charged atoms
pairs $O_{W}$ and $N$. Perhaps the most intriguing feature is the
fact that the short range correlations ($r<10\mathring{A}$) have
amplitudes increasing with amine concentrations, and for all three
$g(r)$ functions, while in the panels below, the pre-peaks and anti-peaks
of the structure factors show the opposite behaviour. This is because
these pre-peaks reflect the large $r$ domain correlations, which
are farther long ranged as the amine concentration \emph{decreases}. 

These findings are in line with those observed in the snapshots of
Fig. \ref{Fig.snap-conc}: the domain correlations start at larger
distances for the smaller amine concentration, since the water domains
are larger. Finally, we see that the like atom correlations produce
positive pre-peaks, while the cross atom correlations (middle panel)
produce negative anti-peaks. As the water concentration becomes smaller,
the amine correlation functions tend to look more like those of neat
amine \cite{2024_Neat_Amines}.

\subsection{Temperature dependence}

\begin{figure}
\centering
\includegraphics[scale=0.4]{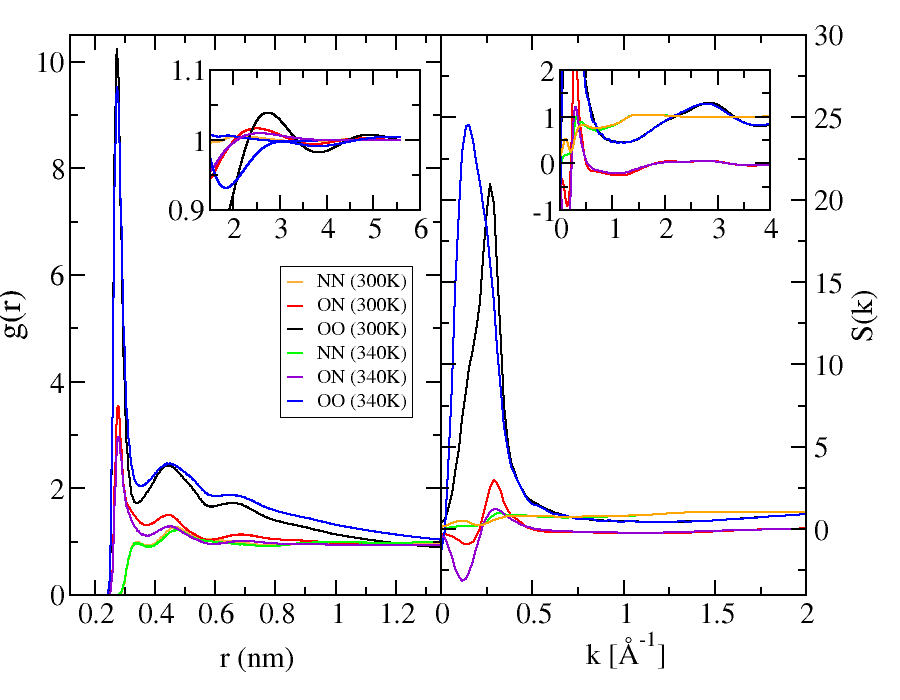}

\caption{Temperature dependence of the pair correlation functions (left panel)
and corresponding structure factors (right panel) for the 20\% aqueous
hexylamine mixtures. The curves for 300 K are shown in thin lines
and those for the higher 340 K are shown in thicker lines. The insets
show the large $r$ (left panel) and small $k$ (right panel) behaviour.}

\label{FigT-dep}
\end{figure}

The temperature dependence of the correlations is illustrated for
the case of hexylamine in Fig. \ref{FigT-dep} and for two temperatures,
namely room temperature $T=300$ K and higher temperature $T=340$
K. Short range correlations ($r<5\mathring{A}$) have more amplitude
at lower temperature than at the higher one, which is a trend generally
observed for simple liquids, since lower temperature means denser
systems and increased correlations. However, at medium distances ($5\,\mathring{A}<r<10\,\mathring{A}$)
we observe a clear increase of correlations. At even greater distances
(left inset), we observe a small shift of the domain correlation oscillations
towards larger distances, indicating that the water domains are slightly
larger at high $T$. The net result is a larger water pre-peak shifted
towards smaller $k$ values, witnessing the larger water domains. 

The temperature dependence is generally less important than the concentration
dependence. Nevertheless, our simulation results are consistent with
the experimental phase diagram (see Fig.SI\_1 in the SI document)
which indicates that the U-shaped 2-phase region (with the LCST at
the minimum of the ``U'') is widening at large temperatures, supporting
the observation that the water domains must be larger. This is equally
supported by the x-ray experiments through the analysis of the SPP
behaviour with temperature, as reported in our recent work\cite{our_expt_amin}.
Indeed, if the water domain increase in size due to higher temperature
thermal dilatation, then, at fixed amine concentration it will become
harder to cover the larger surface, and demixing will occur until
larger amine concentrations that at lower temperature, hence explaining
the U shape curve.

\section{Model dependence}

So far we have examined SPC/E water with CHARMM-AA amines, except
for propylamine where we have used the GROMOS force field model of
our previous simulation results from \cite{2017_Propylamine1,2019_Propylamine2}.
In our investigation we have first tried the GROMOS model, which is
a UA model, and we found that it was demixing at much higher concentrations
than the experimental findings. We then explored OPLS-UA which equally
showed spurious demixing, then OPLS-AA which seemed better, and finally
settled for CHARMM-AA, which was both qualitatively and quantitatively
better. Then we tried the water TIP4P-2005 model, which is generally
considered as superior to the SPC class and TIP class models. Surprisingly,
we found that this water model was not so good for CHARMM-AA amines.
Since our goal was not to test all possible combinations, but rather
find a good compromise in order to analyze the local micro-structure,
we finally settled for the SPC/E-CHARMM-AA combination.

\subsection{Water model}

\begin{figure}
\centering
\includegraphics[scale=0.4]{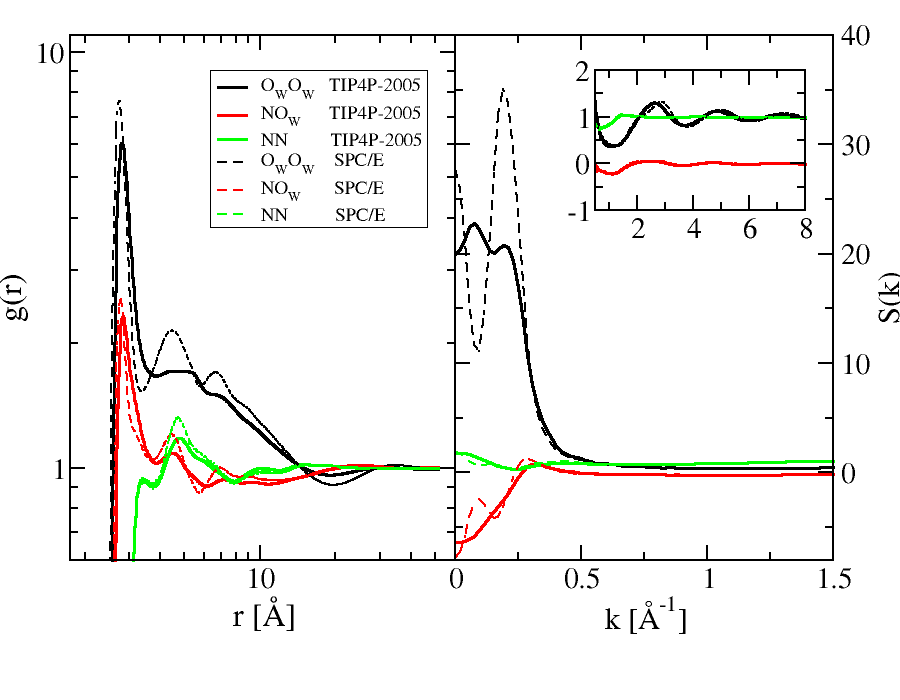}

\caption{Water model dependence of the water oxygen atom O\_W and nitrogen
atom N for the case of the 20\% aqueous hexylamine. The data for the
TIP4P-2005 is shown in thick lines and that for the SPC/E model in
dashed lines. The color conventions are in the inset of the left panel,
which shows g(r) function. The right panel shows the S(k) functions
with focus on small-k part in the main panel and large k part in the
inset.}

\label{Fig-TIP}
\end{figure}

Fig. \ref{Fig-TIP} shows the charged site correlations comparing
the CHARMM-AA 20\% hexylamine model mixed with TIP4P-2005 water model
(thick lines) and the SPC/E water model (dashed lines). It can be
seen that the global trends are generally the same, with perhaps less
structured water correlations in the medium distance range, as can
be seen by the weaker second and third neighbour correlations of $g_{O_{W}O_{W}}(r)$
(black curves). 

Also the amplitudes of the domain oscillations are somewhat smaller.
These features translate into a much less well defined pre-peak for
the water-water structure factor, which looks more like a k=0 concentration
fluctuation peak. The lack of short range structure translates also
in a less pronounced split main peak in the thicker black curve (right
inset).

\subsection{Solute models}

Together with SPC/E water, we have studied four alkylamine models,
namely two united atom models (where the alkyl tail methyl groups
are represented as a single atom), namely OPLS-UA, GROMOS-UA, and
two all atom model, namely OPLS-AA and CHARMM-AA. In a previous study
\cite{2019_Propylamine2} we have studied the GROMOS model for aqueous
propylamine and found it rather good enough to reproduce x-ray and
neutron scattering data trends properly, and superior to the OPLS-UA
model. However, when extending the GROMOS model to longer alkyl tails,
we found that it would show clear demixing for aqueous hexylamine
and octylamine at ambient conditions and 20\% amine content, which
is variance with the experimental data \cite{Phase_diagram_Glinski1994}.
Even though mixing was recovered at higher amine contents, we discarded
these two models, and switched to AA models. Despite both AA models
showing a good mixing trends, it was found that CHARMM-AA was in closer
agreement with experimental X-ray scattering data, the OPLS-AA tending
to overestimate the scattering pre-peak by showing stronger water-amine
head group dimerising that the CHARMM-AA model.

\begin{figure}
\centering
\includegraphics[scale=0.4]{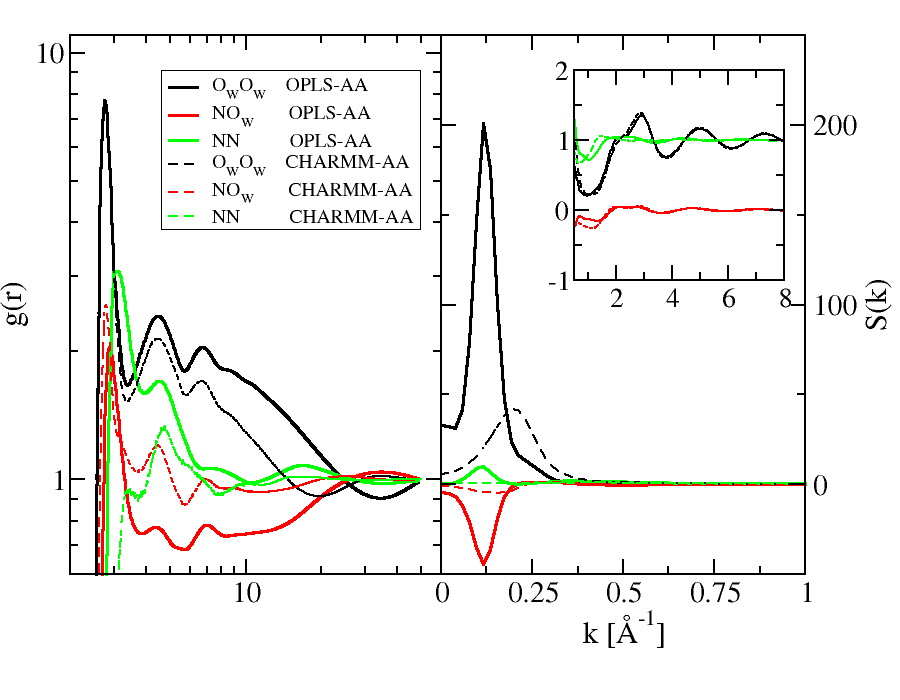}

\caption{Alkylamine model dependence of the water oxygen atom O\_W and nitrogen
atom N for the case of the 20\% aqueous hexylamine. The data for the
OPLS-AA model of the amine is shown in thick lines and that for the
CHARMM-AA model in dashed lines. The color conventions are in the
inset of the left panel, which shows g(r) function. The right panel
shows the S(k) functions with focus on small-k part in the main panel
and large k part in the inset.}

\label{FigOPLS}
\end{figure}

Fig. \ref{FigOPLS} shows the 20\% hexylamine aqueous mixtures correlations,
comparing the OPLS-AA (thick lines) with the CHARMM-AA (dashed lines)
and with SPC/E water as common solvent. OPLS-AA tends to have much
better defined nitrogen atom coverage of the water domain surface.
This is surprising because the partial charge on the nitrogen atom
of the OPLS-AA model $q_{N}=-0.9$ is smaller than $q_{N}=-0.968$
for the CHARMM-AA model. However, the OPLS-AA model has no charges
on the H and C atoms of the alkyl tail, except for the first ones. 

We suggest that the strong $O_{W}-N$ binding in the OPLS-AA model
is enforced through the neutrality of the alkyl tails. Whether this
is a genuine physical effect remains debatable.

\section{X-ray scattering\protect\label{sec:X-ray-scattering}}

The experimental x-ray scattering pre-peaks are in fact the result
of all the contributions from atom-atom structure factors, with some
cancellations between the positive pre-peaks and negative anti-peaks.
At present, it is not possible to predict the SPP for a given system
without examining all atom pair contributions. When considering the
scattering intensity $I(k)$, it is convenient to divide it between
neighbouring atom contributions and medium/long range cluster or domain
contributions. The first ones contribute mostly to the main peak of
$I(k)$, whose position $k_{\ensuremath{MP}}$ is generally related
to the mean atom size, usually in the range $k_{MP}\approx1.5-2\mathring{A}^{-1}$,
corresponding to the average atom van der Waals radii $\sigma\approx2\pi/k_{MP}\approx3-4\mathring{A}$.
The pre-peak is related to the mean size of the cluster/domains and
is system dependent, with position $k_{PP}<k_{MP}$ between $k=0$
and $k_{\ensuremath{MP}}$. The connection of $k_{PP}$ with cluster/domain
size is justified through the observation that it is related to the
periodicity of the medium/large $r$ domain oscillations. All atom-atom
structure factors are affected by domain pseudo-periodicity, with
however in phase oscillations between like domains, and counter-phase
oscillations for cross domain correlations. The first ones contribute
positively to the concerned atom-atom structure factors, while the
latter contribute destructively with negative anti-peaks. 

The most interesting aspect is the fact that, since the segregated
domains contain the same molecular species (or those which mix well
within a given domain), all corresponding atom-atom structure factors
have more or less the same pre-peak features which are independent
of the species atomic details, reflecting mostly domain characteristics.
As a result, following Eqs.(\ref{gr-sep},\ref{Sk-sep}) and the Debye
formula Eq.(\ref{Ik}), one can write the radiation scattering intensity
$I(k)$ as a pre-peak contribution $I_{PP}(k)$ plus a bulk one $\Delta I(k)$
\begin{equation}
I(k)=I_{PP}(k)+\Delta I(k)\label{Ik-sep}
\end{equation}
where the SPP contribution $I_{PP}(k)$ of a binary mixture can be
written solely as a species-species contributions. For the aqueous
amines, one would then have, with obvious notations:
\begin{equation}
I_{PP}(k)=n_{W}^{2}I_{WW}(k)+2n_{W}n_{A}I_{WA}(k)+n_{A}^{2}I_{AA}(k)\label{IPP}
\end{equation}
where $n_{w}=3$ is the number of atoms for water and $n_{A}$ is
the number of atoms of the alkylamine molecule.

\begin{figure}[H]
\centering
\includegraphics[scale=0.4]{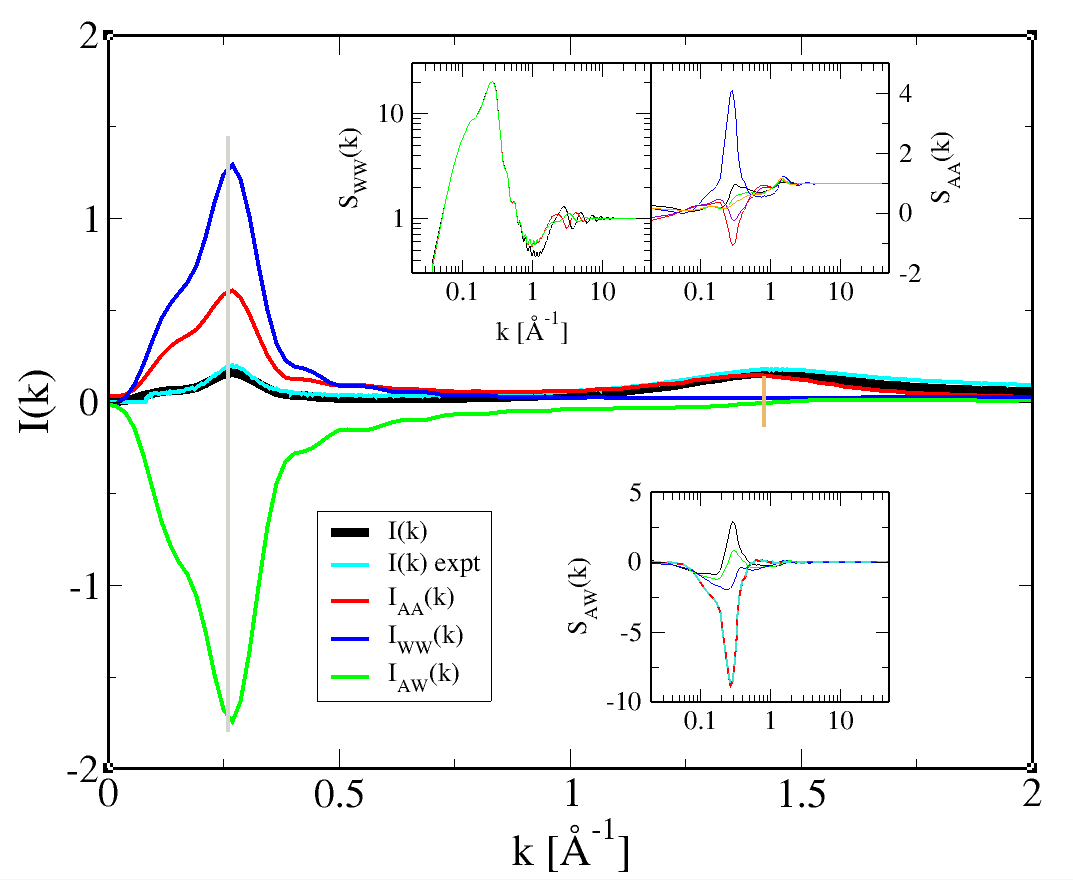}

\caption{Scattering intensity $I(k$) and various contributions (see text)
for the case of the 30\% aqueous hexylamine mixture. The experimental
$I(k)$ is shown in cyan in the main panel and the calculated data
in thick black. The insets show various atom-atom structure factors
$S_{i_{a}j_{b}}(k)$ (see text for details). The upper left inset
is for water $S(k)$ and the upper right inset for hexylamine $S(k)$.
The lower inset is for cross species $S(k)$.}

\label{Fig.Ik}
\end{figure}

Fig. \ref{Fig.Ik} illustrates these arguments for the case of 30\%
aqueous hexylamine. This concentration is not too close to the phase
separation region, as not to be affected by concentration fluctuations,
while still showing appreciable micro-segregation. 

The calculated (black) and the experimental (cyan) $I(k)$ are seen
to be in excellent agreement for this SPC/E-CHARMM-AA model combinations.
The black curve is the result of cancellations between the positive
pre-peak contributions $I_{WW}$ (blue) and $I_{AA}$ (red) with the
negative cross species anti-peak contribution $I_{AW}$(green). The
gray vertical line indicates the position of the pre-peaks $k_{PP}\approx0.24\mathring{A}^{-1}$
(corresponding to domain sizes of $d=2\pi/k_{PP}\approx25\mathring{A}$)
and the light orange vertical line indicates the main peak position
at $k_{MP}\approx1.42\mathring{A}^{-1}$(corresponding to mean atom
size $\sigma=2\pi/k_{MP}\approx4.2\mathring{A}$, which is roughly
that of the carbon atom in the CHARMM-AA model - since there are many
such atoms, it is natural that the main peak position is dominated
by this contribution). It is important to stress the difference in
amplitude between the observed SPP and the three underlying very large
positive and negative molecular contributions, which are the genuine
contributions, although not observable.

The three insets show some of the atom-atom structure factors for
the water (upper left inset), amine (upper right inset) and cross
species (lower inset). For the water $S(k)$ it is clearly seen that
all the three structure factor pre-peaks are totally superposed, hence
justifying the separation in Eq.\ref{Sk-sep}. The main peak contributions
show differences due to different atom pair contributions. The situation
is not so ideal for the two other insets, since both positive and
negative contributions can be observed. However, it must be stressed
that it was intentionally shown, to demonstrate that differences between
pre-peaks exist in most cases. Nevertheless, the dominant contributions
are from peaks that nearly superimpose, as for instance the two water-$C_{T}$
correlations ($O_{W}-C_{T}$ and $H_{W}-C_{T}$) in the lower panel
(cyan and red curves). 

\section{Discussion\protect\label{sec:Discussion}}

The key finding of this study is the fact that stable micro-heterogeneity
is obtained through the stabilizing of the domains interfaces. Indeed,
this seems to be a necessary condition. In the present case, this
stabilization is obtained through the strong binding of water-nitrogen
pairs. Micro-heterogeneity implies the existence of clustered entities,
and these can come in many dimensions, such as the linear chain-like
aggregates of the hydroxyl head groups in alkanols, or domain surfaces
such as in the present case. However, the presence and stabilization
of such topological supra-molecular shapes necessitates that two antagonist
correlations take place: the correlations between the associated charged
groups and those between the charged and uncharged (or weakly charged)
ones. The spatial positioning of the resulting segregated domains,
which is a combination of alternation of domains with spatial disorder,
is reminiscent of the charge order in ionic melts. We have previously
discussed the analogy between charge order and domain order \cite{AUP_Charge_ordering_prepeak_neat_alc,AupDomainOrdering}.
The most important consequence of these microscopic molecular arrangements
is the fact that the experimental observable, namely the scattering
intensity is in fact a combination of the canceling between these
antagonist correlations. Hence, while it is reasonable to attribute
the presence of an SPP to the existence of supramolecular aggregates,
what is really measured is the small remainder between the enormous
real molecular contributions between the antagonist correlations.

The present approach is based on correlation functions analysis. It
can be compared with ``chemical'' approaches, as spectroscopy, which
can provide only information about short time dynamics and near neighbour
positioning. This approach is likely to miss the domain formation
when those are in the 1nm range. Another ``chemical approach'' based
on the so called ``polar/apolar'' paradigm interprets charge order
in terms of dipole order \cite{Triolo_Russina_Amphiphile,Ionic_polar_apolar_Kayshap}.
Since dipole is one step lower than charge in the electrostatic order,
this approach misses the domain order based on charge order. As stated
in Section \ref{subsec:Duality-of-concentration} the chemical approaches
miss the microscopic role of density and concentration fluctuations
which are fundamental random variables in a statistical approach.
Conversely, the present approach does not provide deeper information
about the water-nitrogen dynamics and kinetics, which are readily
provided by chemical approaches. It is more suitable to analyze equilibrium
structure of microscopic molecular configuration over all ranges.
In addition it is very much appropriate to directly interpret radiation
scattering data.

It should be stressed that existence of clusters or supra-structure
can only be searched by analyzing the various atom-atom structure
factors, which can only be obtained through computer simulations of
with model molecular force fields. At present, there are no ``exact''
force field models, hence an necessary incertitude is always present.
Nevertheless, the very good agreement between calculated scattering
intensities and measured ones supports the quality of the MD simulation
model and helps to find proper force fields that describe the real
systems in the best way. Hence, x-ray scattering experiments are a
good test for the models, and the exceptional agreement reported in
our recent work \cite{our_expt_amin} supports the assumption that
the models describe the real life scenario, since the calculated intensities
are well in line with the experiment in all three aspects, concentration
dependence, chain length dependence and temperature dependence.

\section{Conclusion\protect\label{sec:Conclusion}}

In this work, we have shown, through the analysis of the specific
atom-atom density pair correlations, that the aqueous amine mixtures
exhibit clear micro-heterogeneity due to the fact that the amine nitrogen
atoms tend to saturate the surface of the water domains, hence stabilizing
the nano-domain segregation. This is in stark contrast with the case
of the aqueous alcohol mixtures, where the hydroxyl groups do not
tend to stabilize the putative water domain. Hence, even though the
first evidence of micro-segregation was observed in computer simulations
of aqueous-methanol mixtures \cite{MH_Soper_Dixit2002}, this micro-segregation
is not stable for large alcohols, when the hydrophobic alkyl tails
dominate the kinetics of the phase separation. This is the main difference
between the aqueous amines and aqueous alcohol mixtures. The water
oxygen and amine nitrogen pairing supports the LCST scenario proposed
by Walker and Vause, and complements this scenario by the fundamental
fact that this pairing leads to a stability of the segregated nano-domains.
We conjecture that the stabilization of the water domains by local
tight water-solute binding is the general scenario with all types
of stable micro-segregation, which should also generalize to soft
matter systems. This scenario allows the formation of local ``molecular
topological shapes'' which are at the origin of the scattering pre-peaks.
However, the concentration fluctuations associated with these formations
are likely to shift the pre-peak positions to a $k=0$ peak. This
scenario has been recently reported for mixtures of alcohols and ionic
liquids\cite{ion-alc} , where the $k=0$ raise is often the precursor
signature of phase separation in this case. The verification of this
conjecture through the theoretical approaches and simulations of models
could help better understand the role of micro-heterogeneity in various
systems of physico-chemical and bio-chemical interest.

\section*{Supporting Information}

The Supporting Information document contains various figures complementary
to those of the main paper, as well as tables of the force field charge
parameters.

\section*{Acknowledgments}

We thank DELTA for providing synchrotron radiation at beamline BL2
and technical support. The help of Jaqueline Savelkous, Eric Schneider,
Nicola Thiering and Dirk L\H{u}tzenkirchen-Hecht is thankfully aknowledged.
This work was supported by the BMBF via DAAD (PROCOPE 2024-2025, Project-ID
57704875) within the French-German collaborations PROCOPE (50951YA),
\emph{Analysis of the molecular coherence in the self-assembly process:
experiment and theory}.

\bibliographystyle{jpcb_final.bst}
\bibliography{amines}

\end{document}